\documentclass[preprint,3p,11pt,numbers,sort&compress]{elsarticle}
\usepackage{setspace}  
\usepackage{helvet}    
\usepackage[utf8]{inputenc}  
\usepackage[T1]{fontenc}
\usepackage{amssymb}

\usepackage{graphicx}
\usepackage[version=4]{mhchem}
\usepackage{amsmath}
\usepackage{amssymb}
\usepackage{amsfonts}
\usepackage{color}
\usepackage{bm}
\usepackage{flushend,cuted}
\usepackage{epstopdf}
\usepackage[colorlinks,linkcolor=red,anchorcolor=blue,citecolor=green,CJKbookmarks=True]{hyperref}
\usepackage{tabularx} 
\usepackage{float} 
\usepackage{etoolbox}
\usepackage{caption}
\usepackage{multirow}
\usepackage{multicol}
\usepackage{lipsum}  
\usepackage{tikz}  
\usepackage{subfig}
\usepackage{booktabs}

\usepackage[defaultcolor=red]{changes}  
\newcommand{\red}[1]{\textcolor{red}{#1}}

\newcommand{\beq}{\begin{equation}}
\newcommand{\eeq}{\end{equation}}
\newcommand{\beqn}{\begin{eqnarray}}
\newcommand{\eeqn}{\end{eqnarray}}
\newcommand{\bea}{\begin{array}}
\newcommand{\eea}{\end{array}}
\newcommand{\bsub}{\begin{subequations}}
\newcommand{\esub}{\end{subequations}}
\newcommand{\bpm}{\begin{pmatrix}}
\newcommand{\epm}{\end{pmatrix}}


\renewcommand{\emph}[1]{\red{\bfseries #1}}

\renewcommand{\cdot}{\centerdot}

\renewcommand{\raggedright}{\leftskip=0pt \rightskip=0pt plus 0cm}

\begin{document}

\begin{frontmatter}

\title{Unraveling the Allosteric Mechanism and Mechanical Stability of Partial and Complete Loss-of-Function Mutations in p53 DNA-Binding Domain.}

\author[BNU-BJ]{Han Zhou}
\author[BNU-BJ]{Tao Zhou}
\author[BNU-BJ,BNU-ZH]{Shiwei Yan\corref{cor}}
\ead{yansw@bnu.edu.cn}
\address[BNU-BJ]{School of Physics and Astronomy, Beijing Normal University, Beijing 100875, China}
\address[BNU-ZH]{Faculty of Arts and Sciences, Beijing Normal University at Zhuhai, Zhuhai 519087, China}
\cortext[cor]{Corresponding author at: School of Physics and Astronomy, Beijing Normal University, Beijing 100875, China}

\begin{abstract}
TP53 is the most frequently mutated tumor suppressor gene in human cancers, with mutations primarily in its DNA-binding domain (p53-DBD). Mutations in p53-DBD are categorized into hotspot mutations (resulting in complete loss-of-function) and non-hotspot mutations (inducing partial loss-of-function). However, the allosteric mechanisms underlying non-hotspot mutations remain elusive. Using p53 dimer as models, we constructed p53-WT, non-hotspot p53-E180R, and hotspot p53-R248W dimer-DNA complexes to compare the structural and functional impacts of these two mutation types. Our results reveal that both mutations weaken intramolecular interactions in p53-DBD and enhance structural flexibility. Specifically, E180R perturbs dimer interface interactions, impairing dimer stability and cooperative DNA binding; R248W disrupts interactions between the L3/L1 loops and DNA, leading to the loss of DNA-binding capacity. Steered molecular dynamics (SMD) simulations further confirm that both mutations accelerate p53 dimer dissociation, with E180R exerting the most prominent disruptive effect on the mechanical stability of the dimer interface.
\end{abstract}


\end{frontmatter}


\section{\textbf{Introduction}}
\begin{figure}[H]
\centering
    \subfloat{\includegraphics[width=0.85\linewidth]{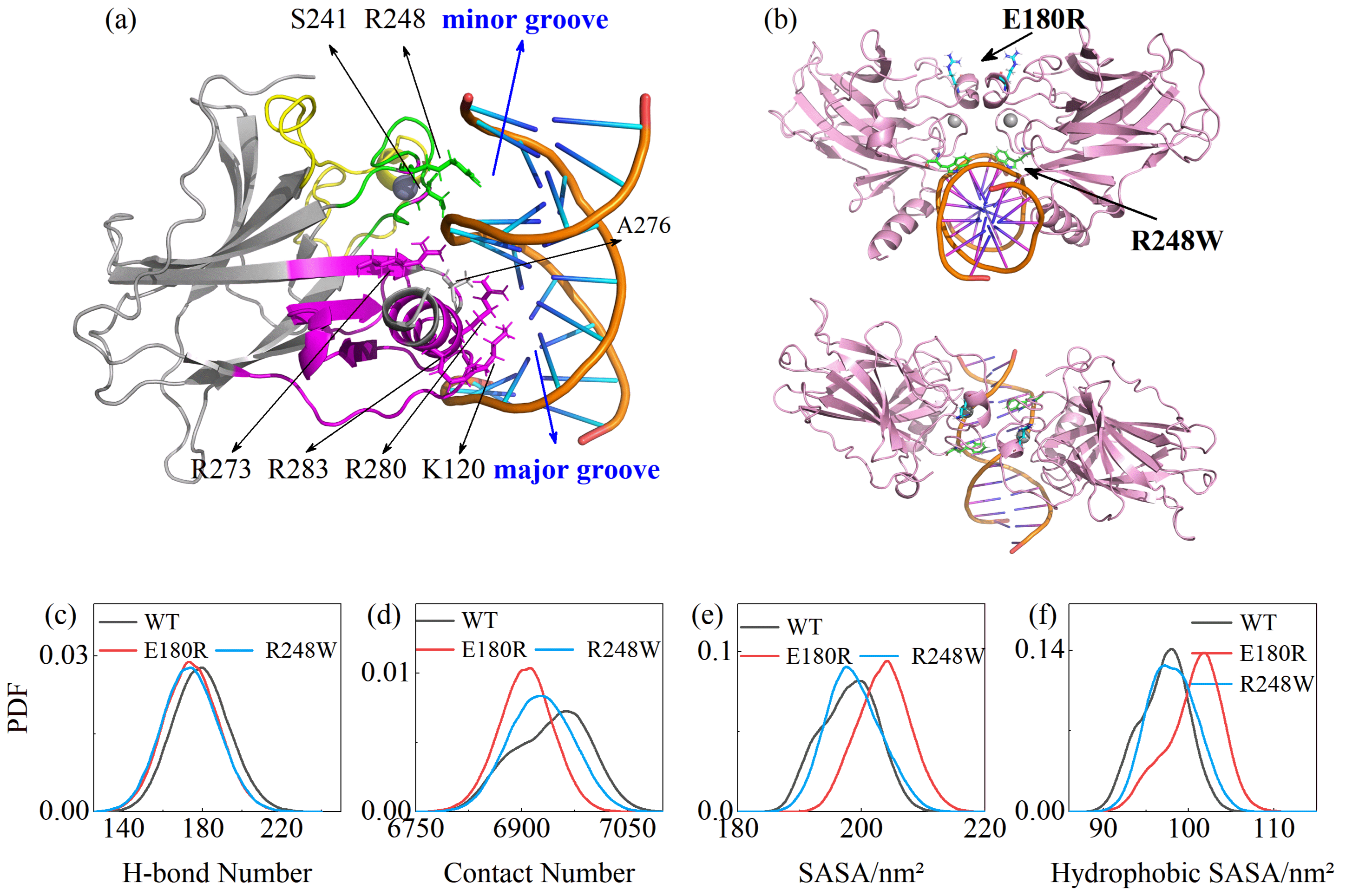}}\\
    \caption{(a) Crystal structure of the p53-DBD in complex with a DNA duplex. The p53-DNA interaction interface is highlighted, including the L3 loop (green), L2 loop (yellow), and loop-sheet-helix (LSH) motif (magenta). Key DNA contact residues for are shown as sticks. (b) Initial structures of the p53-E180R and p53-R248W mutant dimers. (c,d) Probability density distributions of (c) backbone hydrogen bond counts and (d) backbone atomic contact counts in p53 dimers. (e,f) Probability density distributions of (e) solvent-accessible surface area (SASA) and (f) hydrophobic SASA of p53 dimers.}\label{fig-1}
\end{figure}

The tumor suppressor protein p53 is a transcription
factor that regulates the cell cycle, DNA damage repair, and apoptosis, and acts as the ``guardian of the genome''. Mutation-driven loss of function in p53 is closely associated with the development of over 50\% of human cancers \cite{kastenhuber2017putting,levine2020p53,mantovani2019mutant,levine2009first}.  
In human malignancies, p53 mutations are predominantly localized in the DNA-binding domain (p53-DBD), with 30\% targeting mutational hotspot residues \cite{cho1994crystal,joerger2008structural,palanikumar2021protein}. Such hotspot mutations induce largely loss of transcriptional activity either by disrupting DNA-binding interactions (e.g., R248W, R273H) or destabilizing the p53 protein (e.g., R175H, R249S) \cite{milner1991cotranslation,baugh2018there,giacomelli2018mutational}.  The remaining 70\% of non-hotspot mutations typically lead to partial loss-of-function (partial LOF) with residual transcriptional activity \cite{kato2003understanding}. However, the allosteric mechanisms mediated by non-hotspot mutations and the therapeutic strategies for their corresponding partial-LOF phenotypes remain largely elusive.

p53-DBD adopts a $\beta$-sandwich as its core scaffold, supporting two large loops (L2, L3 loop) and a loop-sheet-helix (LSH) motif (Fig. \ref{fig-1}a). The LSH motif and L3 loop form the DNA-binding interface of p53: residues from the LSH motif interact with the major groove of DNA, while residues from the L3 loop engage with the minor groove \cite{cho1994crystal,ho2006structure,kitayner2006structural}.
Under physiological conditions, p53-DBD assembles into a functional tetramer to exert its transcriptional regulatory function, with dimer formation serving as a critical intermediate step in tetramer assembly. Specifically, p53-DBD first forms a stable dimer that binds DNA cooperatively, and two such dimers subsequently assemble into a tetramer via dimer-dimer interactions. In our previous studies, we systematically elucidated the assembly mode of the p53 tetramer and the core intersubunit interaction network \cite{zhou2024mechanisms,zhou2024deciphering}. We clarified the critical role of intermolecular interactions mediated by the H1 helix region (P177-R181) in maintaining the conformational stability of the dimer and ensuring the cooperative p53-DNA binding. Furthermore, we unraveled the molecular mechanisms of mutation-mediated allosteric effects and dominant negative effects in p53 heterotetramers under pathological conditions \cite{zhou2025allosteric}. Based on these studies, we herein investigate the specific pathogenic mechanisms of different types of p53 mutations (especially hotspot and non-hotspot mutations) and the development of precise targeted therapeutic strategies.

Recent studies have identified a class of non-hotspot mutations (e.g., the E180X family mutations) \cite{dehner2005cooperative,klimovich2022partial}. These mutations are frequently detected in germline p53 gene mutations associated with Li-Fraumeni syndrome (LFS). LFS is a familial, dominantly inherited cancer characterized by a wide range of neoplasms occurring in children and young adults \cite{varley2003germline}. The E180X mutations, also referred to as cooperative mutants, promote tumorigenesis by impairing the cooperative DNA-binding capacity of p53. Compared with the complete LOF hotspot mutations (such as R248W, R175H), E180X mutants retain partial transcriptional activity in p53-deficient cancer cells \cite{klimovich2022p53}, suggesting distinct allosteric regulatory mechanisms between these two mutation types.

To investigate the allosteric regulatory mechanisms of different mutation types, we constructed three p53 dimer-DNA complex systems: p53-WT (wild-type p53), p53-E180R (a non-hotspot/partial LOF mutation), and p53-R248W (a hotspot/complete LOF mutation). E180R is a charge-reversal mutation that does not belong to single-nucleotide variants and has therefore not been found in cancer patients so far. However, compared with other cooperative mutants, it exhibits more well-defined mechanistic characteristics regarding protein structural maintenance, cooperative DNA binding, and target gene activation, making it an ideal model for studying cooperative mutants \cite{klimovich2022p53}. The p53 dimer was selected as the model system based on two key considerations: first, previous studies have confirmed that the p53 dimer is a co-translationally formed stable functional unit, whose structural dynamic characteristics and inter-dimer interaction patterns are highly consistent with those of the tetramer  \cite{nicholls2002biogenesis,zhou2024mechanisms}; second, the E180R mutation site is located at the p53 dimerization interface, and the use of a dimer model allows for precise elucidation of the regulatory effects of this mutation on p53 structure and function. Therefore, the dimer model can accurately reveal the impact of mutations on p53 structural and functional regulation.

\section{Methods}

\subsection{System Preparations and MD Simulations Setup}
The initial structure of the p53-WT dimer was taken from a self-assembled p53 tetramer (PDB: 3KMD) \cite{chen2010crystal}. This crystal structure contains four p53-DBD monomers and two DNA half-sites; we extracted two p53 monomers (chains A and B) and a bound DNA half-site as the initial state for our simulations. The p53-E180R and p53-R248W mutant dimers were generated by site-directed mutagenesis of chains A and B using PyMOL software. The initial structures of the p53 dimer systems are shown in Fig. 1b. All-atom molecular dynamics (MD) simulations were performed using the GROMACS 2018 software package \cite{abraham2015gromacs}. The ff14SB force field and OL15 force field were employed for the protein and DNA components, respectively \cite{maier2015ff14sb,zgarbova2015refinement}. The zinc ion in the p53-DBD, which is coordinated by four residues, was parameterized using zinc ion coordination parameters established by Lu et al \cite{lu2007molecular}. All systems were solvated in a dodecahedral box of TIP3P water molecules supplemented with 0.15 mol/L NaCl. Periodic boundary conditions were applied, with a minimum distance of 1.2 nm maintained between the protein and the box edges.

To initiate the simulations, each system underwent sequential equilibration steps: Energy minimization was conducted in two stages (steepest descent followed by conjugate gradient) until the maximum force of the system dropped below 100 kJ$\cdot$ mol$^{ -1}\cdot$ nm$^{ -2}$. The energy-minimized systems were equilibrated under NVT and NPT ensembles sequentially, with the simulation temperature and pressure set to 310 K and 1 bar, respectively. Electrostatic interactions were calculated using the Particle Mesh Ewald (PME) method with a real-space cutoff of 0.8 nm; van der Waals interactions were computed with a 0.8 nm cutoff, and the Verlet buffer list was used for neighbor list updates \cite{pall2013flexible,darden1993particle}. For each system, three independent 800-ns simulations (MD1, MD2, MD3) were performed, yielding a cumulative simulation time of 7.2$\mu s$. The simulation time step was set to 2 fs, and trajectory coordinates were saved every 100 ps.

\subsection{Steered molecular dynamics (SMD) simulations}

Constant-velocity steered molecular dynamics (SMD) simulations were performed to investigate the mechanical stability of the p53-WT, p53-E180R, and p53-R248W dimers \cite{izrailev1999steered,isralewitz2001steered}. 
The final conformational frame from each all-atom MD simulation was used as the initial structure for SMD simulations. To ensure sufficient dissociation, the simulation box was extended to $16 \times 16 \times 16$ nm$^{3}$ and solvated with TIP3P water molecules. Energy minimization was performed, followed by sequential equilibration under NVT and NPT ensembles with parameter settings consistent with those used in the preceding MD simulations. In SMD simulations, a time-dependent external force was applied to drive the separation of two groups: DNA and partial residues of p53 (chain A) were fixed, while partial residues of p53 (chain B) were pulled. Based on previous studies \cite{mehrafrooz2018mechanical,sedighpour2025computational}, the virtual spring constant and pulling rate were set to 4184 kJ/mol/n{m$^{ 2}$} and 0.01 nm/ps, respectively. The virtual spring constant was chosen to be sufficiently large to transfer almost the entire displacement to p53, and the pulling rate enabled quantification of differences in the maximum separation force and mechanical stability of WT and mutant p53 dimers under rapid tensile loading. The effectiveness of these simulation parameters has been fully verified in determining mechanical properties via SMD simulations \cite{mehrafrooz2018mechanical,sedighpour2025computational}.
For statistical validity, results from three independent SMD simulations were averaged.

\subsection{Analysis Method}
Essential dynamics analysis of simulation data was performed using modules from the GROMACS package. The root-mean-square deviation (RMSD), root-mean-square fluctuation (RMSF), and solvent-accessible surface area (SASA) of the system were calculated using the ``rms'', ``rmsf'', and ``sasa'' tools implemented in GROMACS. Atomic contacts were defined by a distance $\leq$ 0.54 nm for aliphatic carbon pairs and $\leq$ 0.46 nm for all other atom pairs. Calculation of residue contact numbers and generation of contact maps were performed using the program developed by Tang \cite{tang2021unraveling}. Hydrogen bonds were defined by a donor--acceptor distance $\leq$ 3.5 \AA\ and a ${\text{X-H}} \cdots {\text{Y}}$ angle  $\leq$ 30$^{\circ}$ \cite{arunan2011definition}. Hydrogen bond occupancy was defined as the probability that a hydrogen bond forms over the duration of the analyzed trajectory. Salt bridges were identified when the minimum distance between the side-chain COO$^-$ group of negatively charged aspartic acid (Asp) or glutamic acid (Glu) and the NH$_{2}^{+}$ or NH$_{3}^{+}$ group of positively charged arginine (Arg) or lysine (Lys) is $<$ 4 \AA. Graphical analyses and visualizations are performed
using the Pymol package \cite{pymol2020PyMOL}. Simulation convergence was verified using RMSD plots (Fig. S1). Statistical analyses were conducted using data extracted from the final 350 ns of each simulation trajectory (i.e., 450-800 ns).

\section{Results and discussion}

\subsection{Mutant p53 Dimers Exhibit Reduced Structural Stability and Increased Loop Flexibility Compared with the p53-WT Dimer}

We first evaluated the structural stability of the p53-WT dimer and two mutant dimers (p53-E180R and p53-R248W) by analyzing key structural parameters: intramolecular backbone hydrogen bond count, backbone atomic contact count, total SASA, and hydrophobic SASA (Fig. \ref{fig-1}(c-f)). Relative to the p53-WT dimer, both mutant dimers showed decreased counts of backbone hydrogen bonds and atomic contacts (Fig. \ref{fig-1}(c,d)), which are two key indicators of protein structural integrity \cite{xu1997hydrogen}. This result indicated that mutations reduce the structural stability of p53 dimers. Notably, the p53-E180R dimer displayed larger total and hydrophobic SASA values than the p53-WT dimer (Fig. \ref{fig-1}(e,f)). Several studies have reported that enhanced protein solvent exposure promotes inappropriate intermolecular interactions, which in turn facilitate misfolding and subsequent aggregation \cite{dobson2003protein,bom2012mutant,pedrote2018aggregation}. Consistent with this mechanism, the increased solvent exposure observed in the p53-E180R dimer suggests a propensity for p53 misfolding and implies that the E180R mutation may induce substantial conformational changes in p53-DBD. In contrast, the p53-R248W dimer exhibited SASA values comparable to those of the p53-WT dimer (Fig. \ref{fig-1}e,f). Collectively, these results demonstrate that both E180R and R248W mutations weaken intramolecular interactions within p53-DBD. As a typical DNA-contact mutation, R248W exerts limited perturbation on the overall conformation of p53, which is consistent with previous experimental findings on p53 mutations \cite{joerger2006structural,joerger2008structural,xu2011gain}.

\begin{figure}[H]
\centering
    \subfloat{\includegraphics[width=0.85\linewidth]{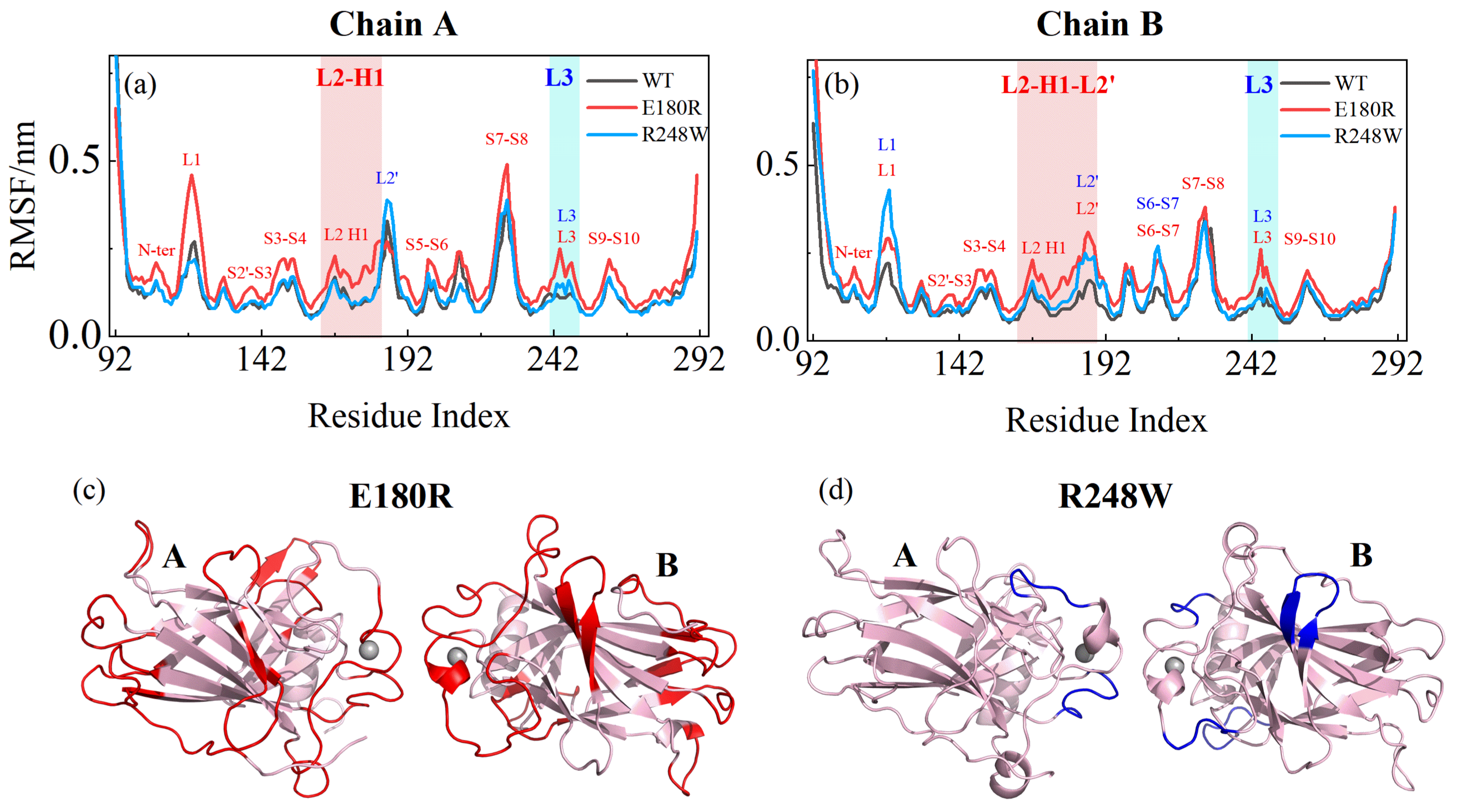}}
    \caption{(a,b) Average root-mean-square-fluctuation (RMSF) of p53 (a) chain A and (b) chain B. Regions with enhanced flexibility relative to the p53-WT dimer are highlighted in the (c) p53-E180R and (d) p53-R248W mutant dimers.}\label{fig-2}
\end{figure}

To investigate the effects of mutations on p53 conformational flexibility, we calculated the RMSF of the $C_{\alpha}$ atoms for the three systems, with the results presented in Fig. \ref{fig-2}(a-b). The p53-DBD contains multiple highly flexible loops (including the L1, L2, and L3 loops) and several $\beta$-turns. Our previous studies have confirmed that these flexible regions play critical roles in specific p53-DNA recognition, dimerization, and dimer-dimer interactions \cite{zhou2024mechanisms,zhou2025allosteric}. Multiple loop regions in the mutant dimers exhibited higher RMSF values than those in the p53-WT dimer. For the p53-E180R dimer, in addition to increase flexibility at the L2 loop (the mutation site), several distal regions, including the L1 loop, L3 loop, and the S2-S2', S3-S4, S5-S6, S6-S7, S7-S8, and S9-S10 $\beta$-turns, also displayed enhanced flexibility (Fig. \ref{fig-2}c). In the p53-R248W dimer, the RMSF of the L3 loop (the mutation site region) was slightly higher than that of the WT, and similarly increased flexibility was observed in the L1 loop, L2' loop, and S6-S7 turn regions (Fig. \ref{fig-2}d). These results indicate that E180R and R248W mutations not only enhance the flexibility of the mutation-containing region but also mediate increased flexibility of multiple loop structures distal to the mutation site. 

\subsection{The E180R Mutation Drives Conformational Rearrangement at the p53 Dimer Interface}

\begin{figure}[H]
\centering
    \subfloat{\includegraphics[width=0.7\linewidth]{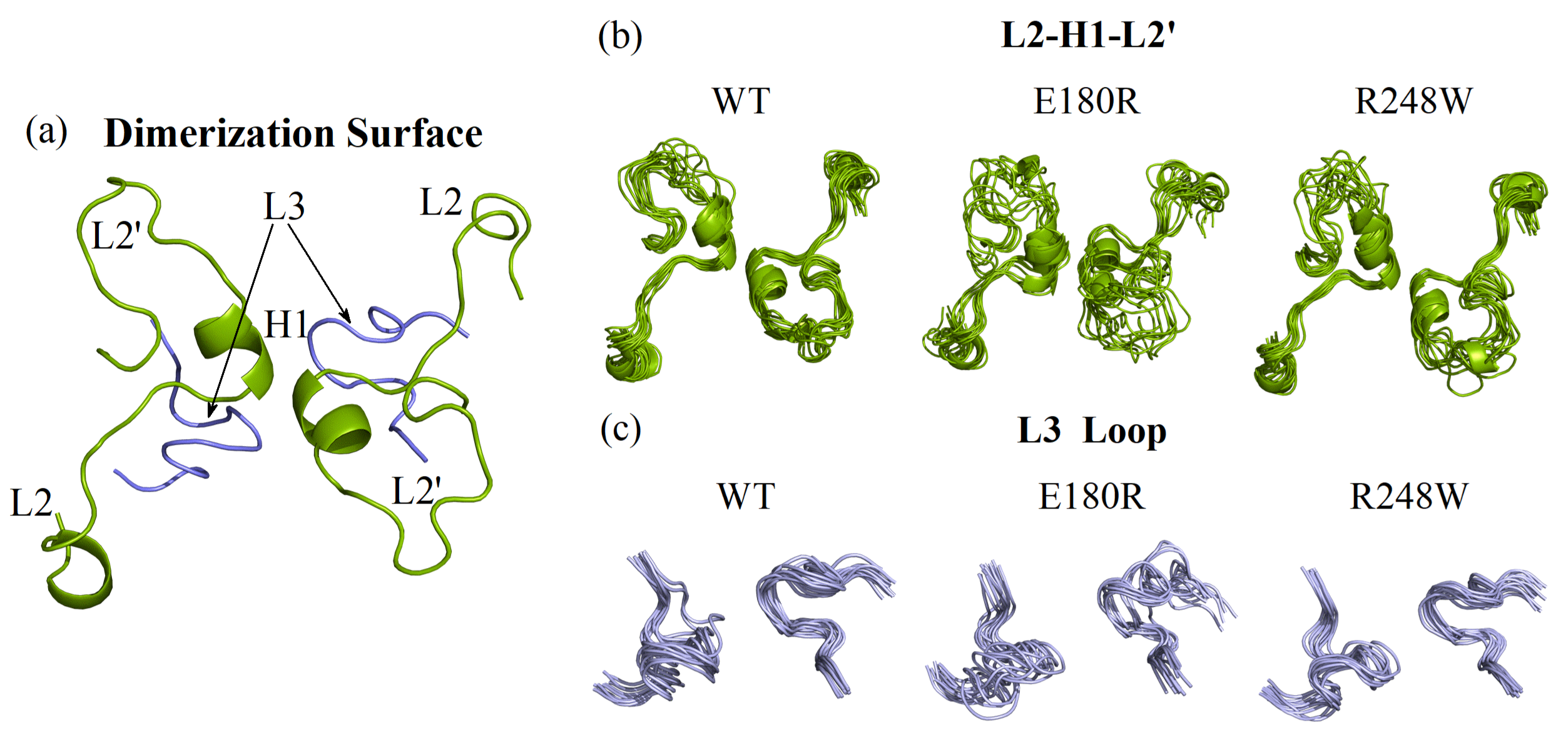}}\\
    \caption{(a) Structure of the p53 dimer interface, comprising the L2-H1-L2' loop and L3 loop. (b, c) Structural alignments of (b) the L2-H1-L2' loop and (c) the L3 loop at 500, 600, 700, and 800 ns frames from three independent simulation replicas.}\label{fig-3}
\end{figure}

Accumulating evidence has demonstrated that the L2 and L3 loops play critical roles in maintaining the structural stability of the p53 dimer interface (Fig. \ref{fig-3}a) \cite{ho2006structure,zhou2024mechanisms}. The L2 loop is a long flexible loop (residues 164-194) with a short helix (H1) embedded within its sequence. Therefore, the L2 loop can be divided into three functional regions: the N-terminal L2 loop (residues 164-176), the H1 helix (residues 177-181), and the C-terminal L2' loop (residues 182-194). To capture the conformational characteristics of the dimer interface, we aligned the structures of the L2 and L3 loops from each system at 500, 600, 700, and 800 ns for three independent replicas, with the results presented in Fig. \ref{fig-3}(b-c). In the p53-WT dimer, the conformations of the L2-H1-L2' loop and the L3 loop are highly conserved. In contrast, the p53-E180R dimer exhibited significantly increased flexibility in both the L2-H1-L2' and L3 loops. For the p53-R248W dimer, only the L2' loop showed increased flexibility, while the conformations of the L2 loop, H1 helix, and L3 loop remained stable.

\begin{figure}[H]
\centering
    \subfloat{\includegraphics[width=0.85\linewidth]{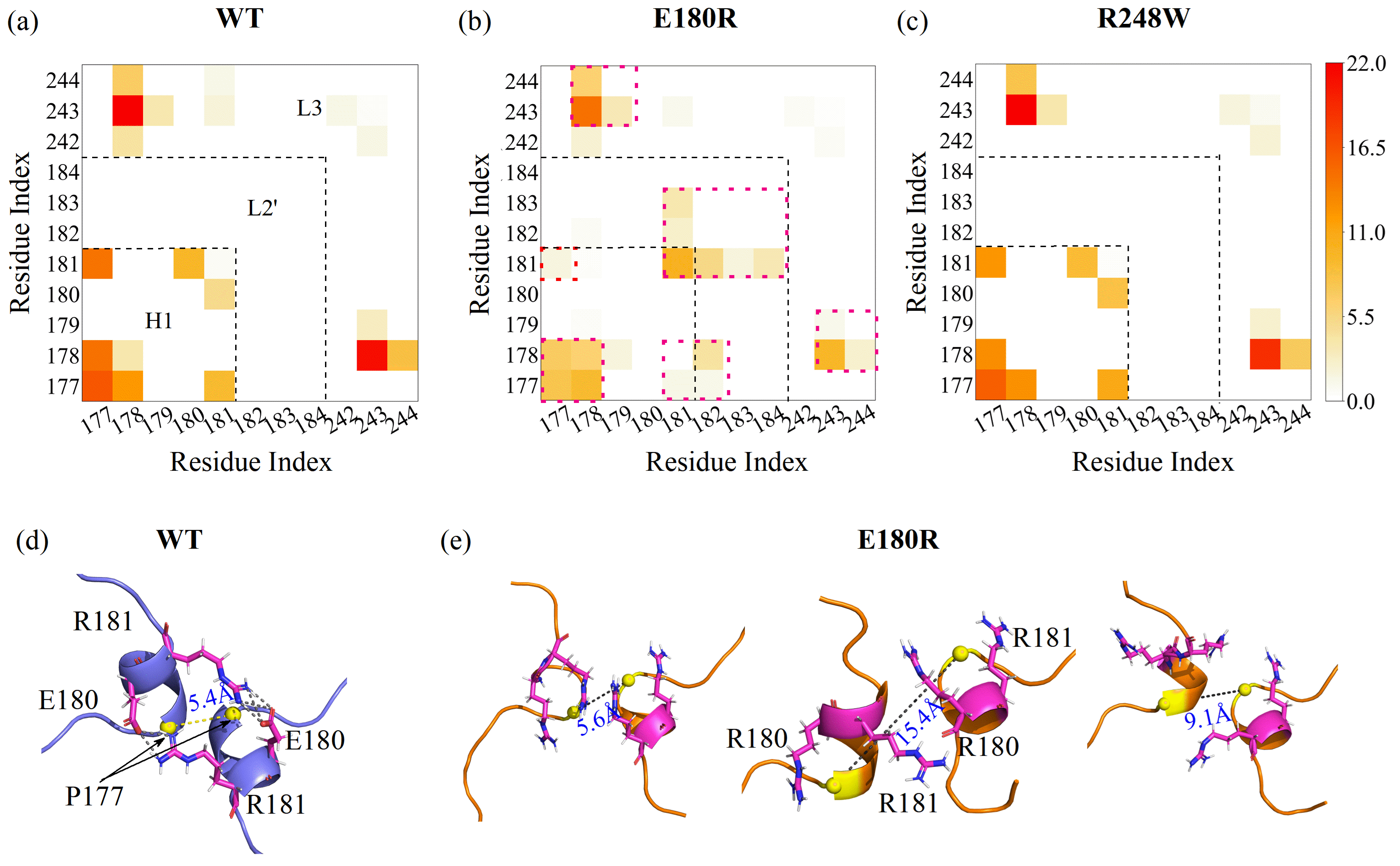}}\\
    \caption{(a-c) Contact maps of the p53 dimer interface in (a) p53-WT, (b) p53-E180R, and (c) p53-R248W dimers.(d, e) Conformations of H1 helix pairs in (d) p53-WT and (e) p53-E180R dimers (for three independent simulation replicas). The $C_{\alpha}$ atoms of residue P177 are represented as yellow sphere.}\label{fig-4}
\end{figure}

At the p53-WT dimer interface, two H1 helices are symmetrically arranged, and two stable salt bridges are formed between residues E180 and R181. These salt bridges constitute the core interactions of the p53 dimer interface \cite{dehner2005cooperative}. However, the E180R charge inversion mutation disrupts charge complementarity at the p53 dimer interface, thereby abolishing the E180-R181 salt bridges. To investigate the allosteric effects of mutations at the dimer interface, we calculated the residue atomic contacts at the dimer interface, with the results presented in Fig. \ref{fig-4}(a-c). For the p53-WT dimer (Fig. \ref{fig-4}a), multiple stable interactions are formed between the H1 helix and L3 loop of the two monomers. In addition to the E180-R181 salt bridges, the H1 helices form multiple pairs of stable residue contacts, including P177-P177, P177-H178, and P177-R181, among others. Stable residue interactions are also formed between the H1 helix and L3 loop, such as H178-M243 and H178-G244. The residue contact pattern at the p53-R248W dimer interface is similar to that of the p53-WT dimer (Fig. \ref{fig-4}c), indicating that the R248W mutation does not perturb the interaction network at the dimer interface. However, the interaction pattern at the p53-E180R dimer interface is significantly altered (Fig. \ref{fig-4}b). Specifically, the contact between R180 (the mutated residue) and R181 is completely lost, while the contact frequency of the R181-R181 residue pair is increased. Additionally, the contact frequencies of P177-P177, P177-H178, and P177-R181, as well as the H1-L3 residue pairs (e.g., H178-M243 and H178-G244), are reduced. Notably, the contact frequencies of the H178-H178 and H178-H179 residue pairs within the H1 helix pair are increased, and R181 forms new contacts with residues in the L2' loop (residues 182-184). These results demonstrate that the conformational changes induced by the E180R mutation alter the interaction network at the p53 dimer interface.

From the structure of the p53 dimer, we identified that P177-P177 is the closest residue pair at the dimer interface. To characterize the spacing changes of the H1 helix pairs, we calculated the distance distribution of the $C_{\alpha}$ atoms of the P177-P177 residue pair for the three systems, with the results presented in Fig. S3. The spacing of the H1 helix pairs remained highly stable in both the p53-WT and p53-R248W dimers (Fig. \ref{fig-4}d), and increased in the p53-E180R dimer. We extracted the most probable conformations of this distance distribution from its three independent simulation trajectories. As shown in Fig. \ref{fig-4}(e), the charge-reversal E180R mutation drives an increase in the spacing between the H1 helix pairs of the two p53 monomers and may also induce a certain degree of rotation of the helix pairs. This conformational change results in a reduction in intermonomer residue contacts, thereby impairing the binding affinity of the p53-E180R dimer. This finding elucidates the core molecular mechanism underlying the reduced cooperative activity of the p53-E180R mutant observed in experimental studies \cite{dehner2005cooperative,klimovich2022partial}.

\subsection{The R248W Mutation Disrupts Interactions Between the L3/L1 Loops and DNA, Leading to Loss of DNA-Binding Capacity}

\begin{figure}[H]
\centering
    \subfloat{\includegraphics[width=0.85\linewidth]{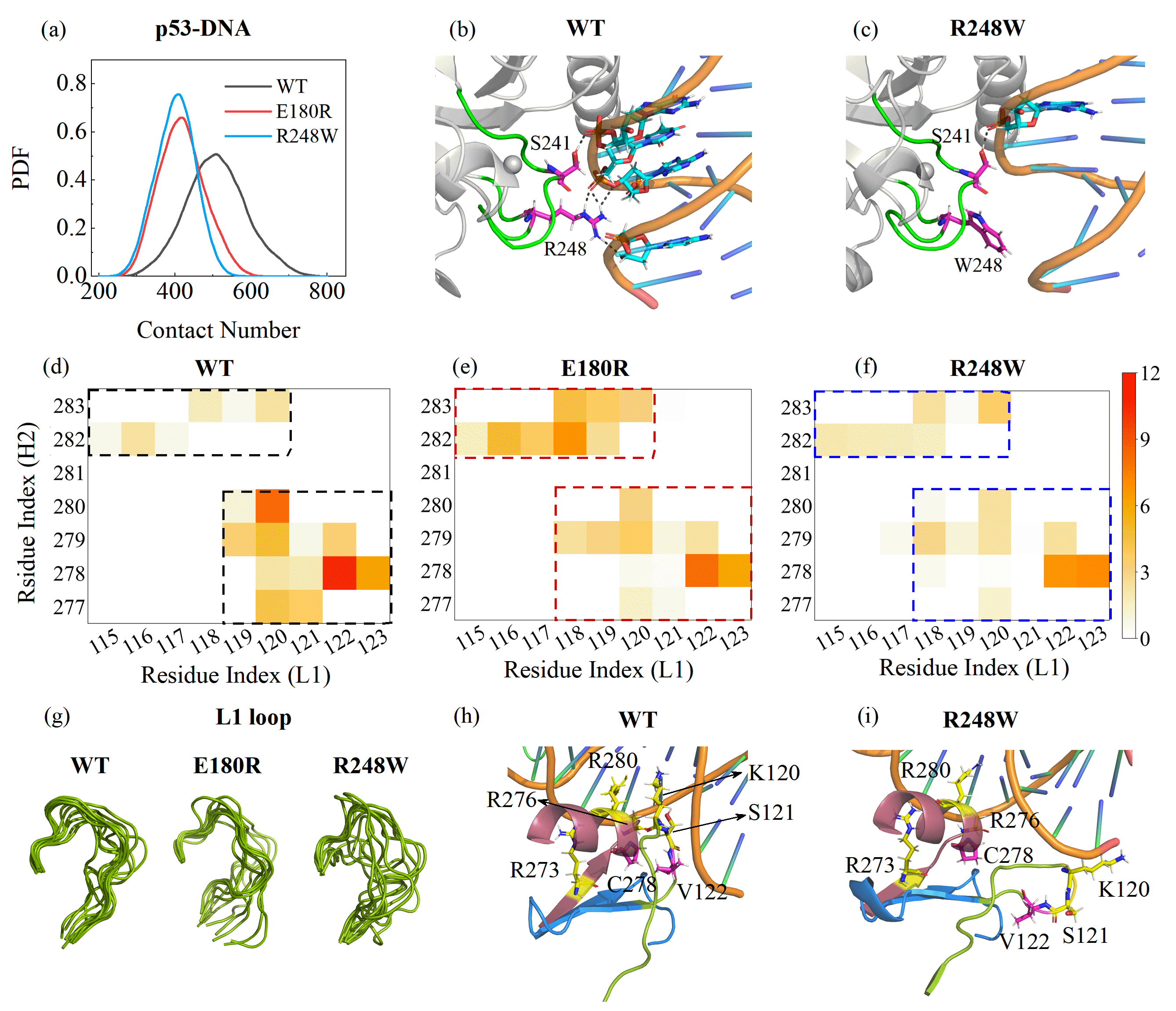}}\\
    \caption{(a) Probability density distributions of the contact numbers between p53 dimer and DNA. (b, c) Contact conformation of L3 loop with the DNA minor groove in the (b) p53-WT dimer and (c) p53-R248W dimer. (d-f)Contact maps of residues in the L1 loop and H2 helix: (d) p53-WT, (e) p53-E180R and (f) p53-R248W dimers. (g) Structural alignments of the L1 loop (chain B) at 500, 600, 700, and 800 ns frames from three independent simulation replicas. The representative structure of loop--sheet--helix motif (chain B) in (h)p53-WT and (i) p53-R248W dimers.}\label{fig-5}
\end{figure}

The transcriptional activation function of p53 is closely related to the specific binding of p53-DBD to DNA \cite{kitayner2006structural}. To investigate the impact of mutations on the DNA-binding ability of p53-DBD, we calculated the total number of contacts between the p53 dimer and DNA in three systems. As shown in Fig. \ref{fig-5}a, both mutant dimers exhibited decreased DNA contact numbers compared with p53-WT, with most probable values of 506.90 (p53-WT), 418.96 (p53-E180R), and 409.12 (p53-R248W), respectively. 
Several studies have confirmed that the L3 loop of p53 binds to the minor groove of DNA, while the LSH motif binds to the major groove \cite{cho1994crystal,ho2006structure,kitayner2006structural}. Thus, the reduced total DNA contacts observed in the mutants imply potential disruptions in these region-specific interactions.

As shown in Fig. S4a, at the interaction interface between the L3 loop and DNA, the p53-E180R dimer exhibited contact levels comparable to those of the p53-WT dimer, indicating that the E180R mutation exerts no effect on the interaction between the L3 loop and the DNA minor groove. However, the R248W mutation site is located in the L3 loop, and the interaction between the L3 loop and DNA was markedly diminished in the p53-R248W dimer. Within the minor groove of DNA, residues S241 and R248 (both in the L3 loop) form stable hydrogen bonds with DNA (Fig. \ref{fig-5}b). We calculated hydrogen bond occupancy to assess binding stability: residue S241 maintained stable hydrogen bond binding capacity in all three p53 systems, with hydrogen bond occupancies for the two subunits of each dimer as follows: WT (82.34\%/89.01\%), E180R mutant (60.21\%/85.89\%), and R248W mutant (90.80\%/96.76\%). In p53-WT and p53-E180R dimers, R248 formed stable hydrogen bonds with DNA; the hydrogen bond occupancies for the two subunits were 56.93\% and 54.64\% for the p53-WT, and 59.72\% and 45.73\% for the p53-E180R.
Notably, the R248W mutation completely abrogated the ability of residue 248 (now W248) to form hydrogen bonds with DNA (Fig. \ref{fig-5}c). However, further structural analysis confirmed that no significant conformational changes occurred in the L3 loop harboring the R248W mutation site; the mutant residue W248 can maintain contacts with DNA via hydrophobic interactions, thereby preserving the structural integrity of the L3 loop.

The H2 helix and L1 loop from the loop-sheet-helix motif fit in the major groove of DNA. As shown in Fig. S4(b-c), the contact levels between the H2 helix and DNA were comparable in three systems. However, the number of contacts between the L1 loop and DNA in the mutant dimers was markedly lower than that in the p53-WT dimer. Such reduction may be associated with the high conformational flexibility of the L1 loop in the mutant dimers (Fig. \ref{fig-2}), and may also impact the conformation and structural stability of the loop-sheet-helix motif. Given that only the L1 loop of chain B in the selected crystal structure can form hydrogen bonds with DNA \cite{chen2010crystal}, this study focused on analyzing the interaction characteristics between the LSH motif of chain B and DNA.

Contact maps of the L1 loop and H2 helix regions showed that in the p53-WT dimer (Fig. \ref{fig-5}d), residues 119-123 of the L1 loop were in close contact with residues 277-280 of the H2 helix (defined as ``Contact 1''), whereas only minimal contact existed between residues 115-120 of the L1 loop and residues 282-283 of the H2 helix (defined as ``Contact 2''). In both mutant dimers, the contact strength of the ``Contact 1'' region was decreased to varying degrees, while that of the ``Contact 2'' region was enhanced to varying degrees (Fig. \ref{fig-5}e-f). These results indicate that mutations can mediate changes in the interaction pattern between the L1 loop and H2 helix within the loop-sheet-helix motif. To clarify the conformational characteristics of the L1 loop, we aligned the L1 loop structures of the three systems at 500, 600, 700, and 800 ns in three independent replicas. As shown in Fig. \ref{fig-5}g, the L1 loops of the p53-E180R and p53-R248W dimers exhibited increased disorder and reduced structural stability.
Hydrogen bond analysis revealed that residues K120 and S121 in the L1 loop form hydrogen bonds with DNA: the hydrogen bond occupancies were 79.54\% (K120) and 67.75\% (S121) in the p53-WT dimer, while decreasing to 56.73\% (K120) and 60.23\% (S121) in the p53-E180R dimer, and to 45.73\% (K120) and 34.14\% (S121) in the p53-R248W dimer. Structural analysis demonstrated that compared with the p53-WT dimer, the high flexibility of the L1 loop in the mutants leads to transient spatial displacement of residues such as K120, S121, and V122. Such displacement makes it difficult to maintain contacts with the H2 helix and the DNA major groove (Fig. \ref{fig-5}h-i). These results indicate that both the E180R and R248W mutations can mediate the rearrangement of the interaction pattern between the L1 loop and the H2 helix, with the R248W mutation largely weakening the stability of hydrogen bonds between the L1 loop and DNA. Crystallography study have reported that the L1 loop of p53-DBD can adopt distinct conformations depending on the binding state of p53 to DNA, which may explain why the L1 loop in the mutants, despite its relatively large spatial distance from the two mutation sites, still exhibits conformational response to the mutations \cite{emamzadah2014reversal}.

\subsection{Steered Molecular Dynamics (SMD) Simulations Reveal the Effects of Two Mutations on p53 Dimer Dissociation}

\begin{figure}[H]
\centering
    \subfloat{\includegraphics[width=0.85\linewidth]{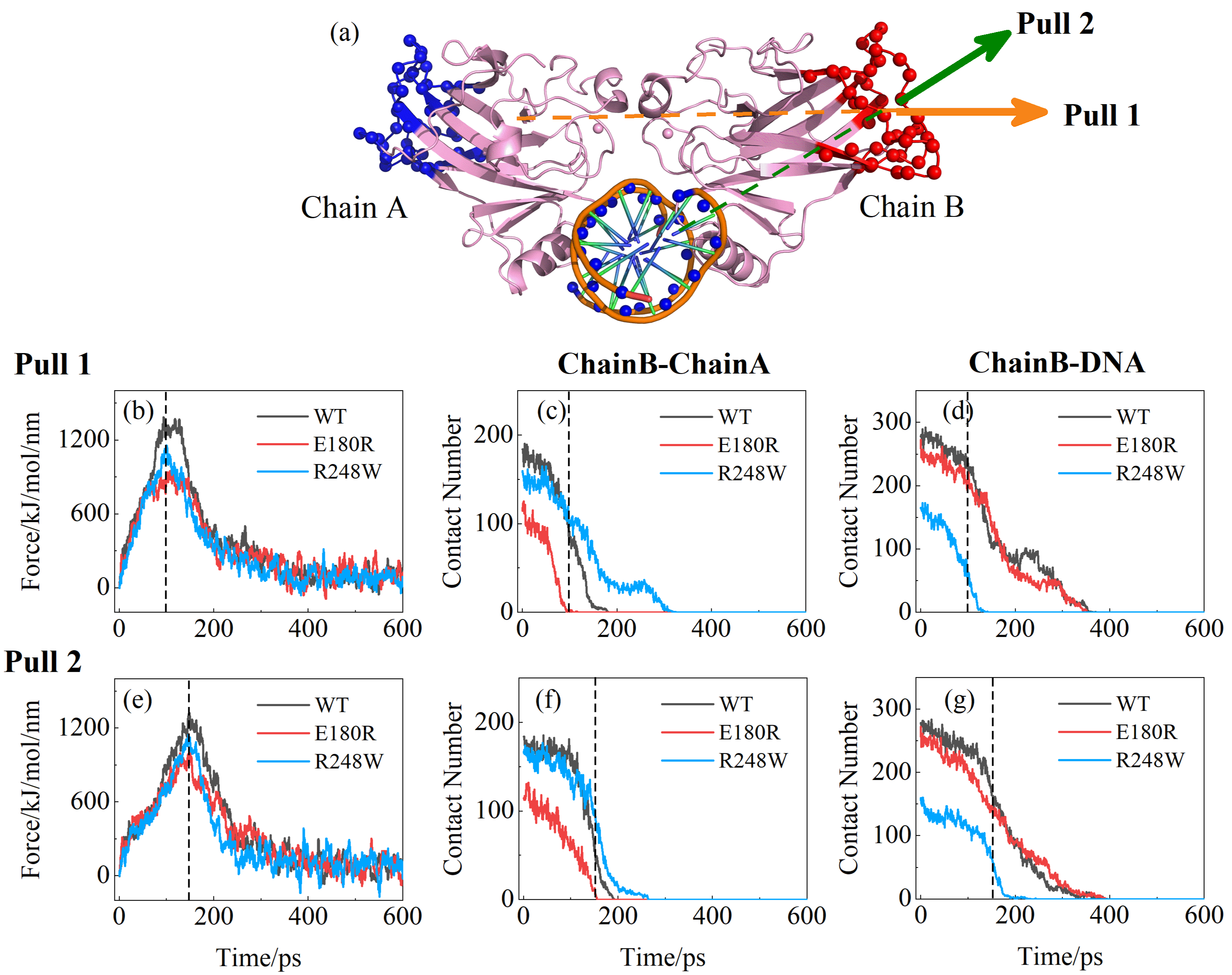}}\\
    \caption{(a) Structure of the p53 dimer complex used for steered molecular dynamics (SMD) simulations. Fixed residue groups are highlighted in blue, and pulling residue groups in red. (b, e) Force-time diagrams of the pulling process for p53-WT, p53-E180R, and p53-R248W dimers under two distinct pulling directions.(c, f) Time evolution of atomic contact numbers between p53 chains A and B under the two pulling directions.(d, g) Time evolution of atomic contact numbers between p53 chain B and DNA under the two pulling directions. The dotted line represents the time frame reaching the maximum force ($F_{max}$).}\label{fig-6}
\end{figure}

To investigate the impact of mutations on the mechanical stability of p53 dimers, we performed SMD simulations to characterize the dissociation dynamics of p53-WT and mutant dimers. As illustrated in Fig. \ref{fig-6}a, the fixed and pulled residues (residues 105--110, 146--156, 218--231, and 257--266) are highlighted in blue and red, respectively. These residues were selected based on interactions between p53 monomers―specifically, the fixed and pulled residues have no dynamical correlation with residues involved in dimerization or p53-DNA binding interfaces \cite{zhou2024mechanisms}. Furthermore, to prevent nonspecific conformational distortion of DNA during the pulling process, we fixed the DNA backbone atoms. To enhance the reliability and replicability of the simulations, we implemented two distinct pulling directions: Pull-1 was applied along the centroid connection between p53 chains A and B, and Pull-2 was applied along the centroid connection between DNA and p53 chain B.

The force-time diagram for the dissociation of the p53-WT dimer and p53-E180R/p53-R248W mutant dimers were shown in Fig. \ref{fig-6}(b,e). 
The maximum force ($F_{max}$) required for dimer unfolding followed the order p53-WT > p53-R248W > p53-E180R, and this trend was consistently reproduced across simulations with the two pulling directions. Force-time diagram for all independent simulation replicas are presented in Fig. S5,S6. 
Notably, the average time required to reach $F_{max}$ in the Pull-1 direction was shorter than that in the Pull-2 direction, indicating that pulling along Pull-1 direction is more effective to p53 dimer dissociation.

To elucidate the molecular mechanisms underlying the differences in $F_{max}$ for three systems, we calculated the atomic contact numbers between the two p53 monomers and between p53 (chain B) and DNA during the pulling process. At the p53 dimer interface (Fig. \ref{fig-6}c,f), the atomic contact numbers decreased progressively with the progression of pulling. Specifically, the intermonomer contact numbers of the p53-WT and p53-R248W dimers underwent a gradual reduction during the force-loading phase, followed by a rapid decline only when $F_{max}$ was attained. The intermonomer contact numbers of the p53-E180R dimer had nearly dropped to zero upon reaching $F_{max}$. Further analysis of the p53-DNA binding interface (Fig. \ref{fig-6}d,g) revealed that the atomic contact numbers of all three systems decreased rapidly after $F_{max}$ was attained, with the p53-R248W dimer requiring the shortest time for complete p53-DNA dissociation. These results demonstrate that the cooperative mutation E180R and DNA-contact mutation R248W accelerate the dissociation process of p53 dimers under mechanical stretching by exerting negative regulatory effects on the p53 dimer interface and DNA-binding interface, respectively. Notably, the p53-E180R dimer required the lowest $F_{max}$, indicating that this mutation exerts the most pronounced disruptive effect on the mechanical stability of the p53 dimer.

\begin{figure}[H]
\centering
\subfloat{\includegraphics[width=0.4\linewidth]{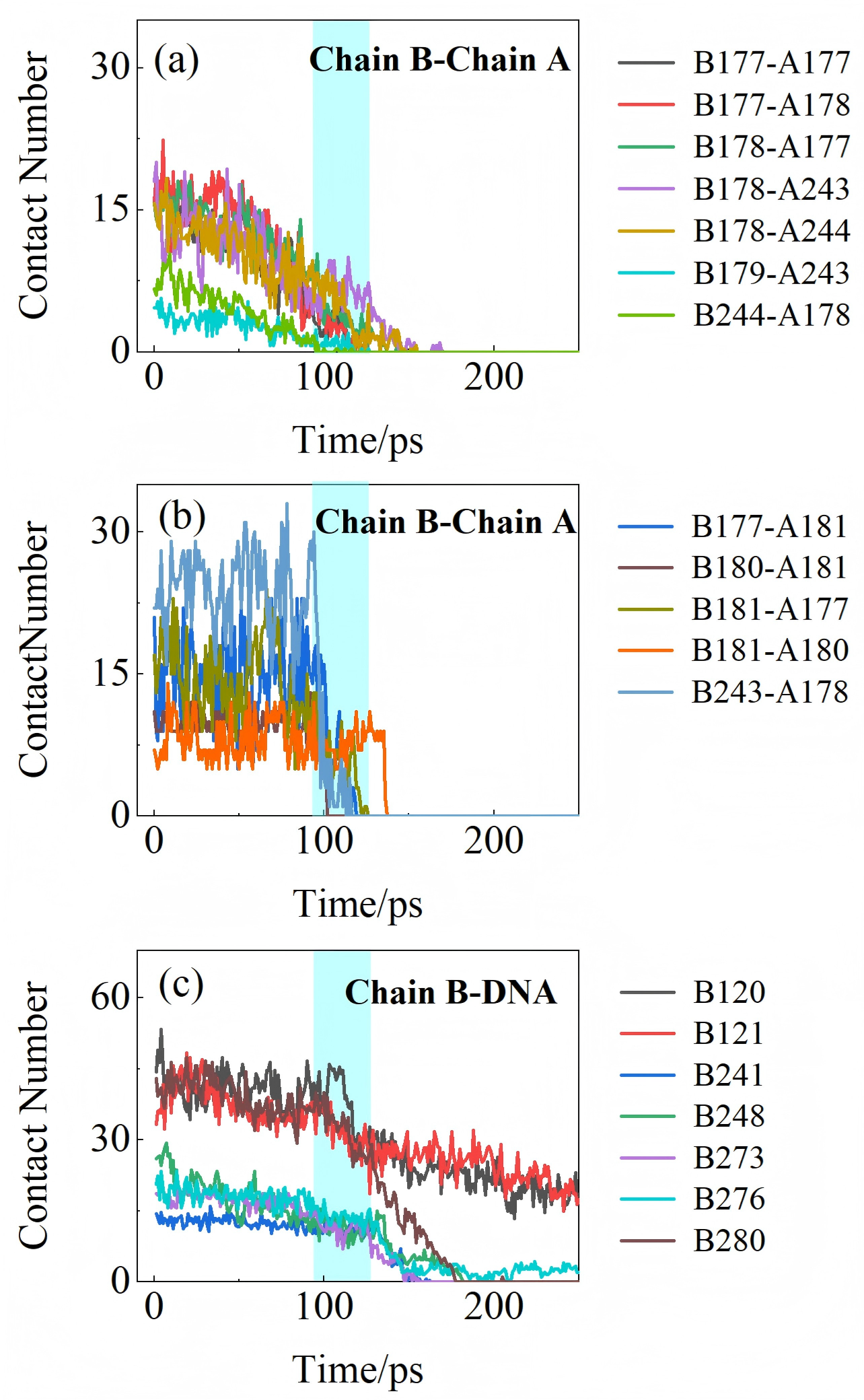}}\\
\caption{(a, b) Time evolution of atomic contact numbers for residue pairs at the dimer interface: (a) residue pairs showing a linear decrease in atomic contact numbers, and (b) those exhibiting high mechanical stability during the pulling process. (c) Time evolution of atomic contact numbers for key p53-DBD residues bound to DNA.}\label{fig-7}
\end{figure}

To evaluate the mechanical stability of residue pairs at each intermolecular interface during the pulling process, we decomposed the total interfacial contacts into individual residue pairs. Fig. \ref{fig-7} presents the results for the p53-WT dimer under the Pull-1 condition, with complete results for the three systems under Pull-1 and Pull-2 conditions provided in Fig. S7 and S8. At the p53 dimer interface, the atomic contact numbers of most residue pairs decreased linearly with the progression of pulling (Fig. \ref{fig-7}a). Notably, the contact numbers of three residue pairs (P177-R181, E180-R181, and M243-H178) remained relatively stable prior to reaching $F_{max}$, followed by a rapid decline immediately after the $F_{max}$ was attained (Fig. \ref{fig-7}b). Therefore, these residue pairs are the core units responsible for maintaining the mechanical stability of the dimer interface, and such stable interactions provide critical structural support for the p53 dimer to resist external mechanical pulling. In the p53-E180R dimer (Fig. S7 and S8), the mutation led to the near-complete disappearance of interactions between the residue pairs P177-R181 and E180-R181, thereby reducing the mechanical stability of the dimer interface.

Furthermore, analysis of the key residues (K120, S121, S241, R248, R273, A276, and R280) at the p53-DNA binding interface showed that their interactions with DNA exhibited a consistent pattern of mechanical stability (Fig. \ref{fig-7}c). During the force-loading phase (before reaching the $F_{max}$), the contact numbers remained stable or decreased slowly, while they dropped rapidly after reaching $F_{max}$. In the p53-R248W dimer, the interactions between W248, K120, S121 and DNA were weakened, resulting in increased susceptibility of the p53-DNA interface to dissociation under mechanical stress. These results indicate that the key interactions at the p53-DNA binding interface exhibit strong mechanical stability, which is crucial for p53 to effectively bind DNA and exert its transcriptional regulatory functions in the mechanical microenvironment.

\section{Conclusion}
In summary, we performed all-atom molecular dynamics (MD) and steered molecular dynamics (SMD) simulations on the dimer systems of p53-WT, non-hotspot/cooperative mutant p53-E180R, and hotspot/DNA-contact mutant p53-R248W. We identify the atomic-level allosteric mechanisms of these two mutation types. As a charge-reversal mutation, E180R induces conformational changes in p53-DBD, specifically the separation and rotation of the H1 helix pair at the dimer interface, thereby reducing p53 dimer affinity. Meanwhile, the E180R mutation exerts minimal effects on the key interactions required for p53-DNA binding, enabling p53-E180R dimer to retain a relatively stable DNA-binding state. This finding may explain why the p53-E180R retains partial transcriptional activation function in p53-deficient cancer cells. The DNA-contact mutation R248W primarily causes loss of DNA-binding capacity while exerting slight impacts on the overall conformation of the p53-DBD. SMD simulation results further revealed the molecular mechanisms through which both mutations accelerate dimer dissociation. We identified key residue pairs with high mechanical stability at the dimer interface and DNA-binding interface; these interactions play crucial roles in maintaining the assembly and stability of the p53 complex within the mechanical microenvironment.

These findings unveil the distinct pathogenic mechanisms of the two mutant types and provide insights for targeted therapeutic strategies. For cooperative mutants, interventions may focus on stabilizing the dimer interface and enhancing the overall structural stability of p53. For DNA-contact mutants, priority should be given to compensating for the loss of DNA-binding capacity, such as inserting binding-enhanced mutations or targeting with small-molecule chaperones \cite{joerger2005structures,sallman2021eprenetapopt}. Some studies have confirmed that tumors harboring cooperative mutations exhibit distinct therapeutic responses compared with those bearing hotspot mutations \cite{timofeev2019residual,klimovich2022p53}. Therefore, when using p53 mutations to predict prognosis or select treatment regimens, cooperative mutations should be distinguished from hotspot mutations.

\vspace{2em}

\bibliographystyle{model1-num-names}
\bibliography{ref}

\end{document}